\documentclass[superscriptaddress,pre,showpacs,showkeys,twocolumn,eqsecnum]{revtex4}
\usepackage{amsmath}
\usepackage{amsfonts}
\usepackage{bm}
\usepackage{graphicx}

\begin{document}

\hyphenation{eigen-values}

\title{
Stiff polymer in monomer ensemble}
\author{K. K. M\"uller-Nedebock}
\email[Author to whom correspondence should be addressed.  E-mail: ]{kkmn@physics.sun.ac.za}
\affiliation{Department of Physics, University of Stellenbosch,
Private Bag X1, Matieland, 7602 South Africa}
\author{H. L. Frisch}
\affiliation{Department of Chemistry, State University of New York at
  Albany, Albany 12222, NY, USA}
\affiliation{Department of Physics, University of Stellenbosch,
Private Bag X1, Matieland, 7602 South Africa}
\author{J. K. Percus}
\affiliation{Courant Institute, 
New York University, New York, 10012,
  NY, USA}
\affiliation{Physics Department, 
New York University, New York, 10012,
  NY, USA}

\date{2002.01.14}

\begin{abstract}
We make use of the previously developed formalism for a monomer
ensemble and include angular dependence of the segments of the
polymer chains thus described.  In particular we
show how to deal with 
stiffness when the polymer chain is confined to certain regions.  
We investigate the stiffness from the perspectives of a
differential equation, integral equations, or recursive relations
for both continuum and lattice models.  Exact analytical solutions are 
presented for two cases, whereas numerical results are shown for a
third case.
\end{abstract}
\pacs{36.20.Ey, 82.35.Lr}
\keywords{Conformation of polymers; Chain stiffness}

\maketitle


\section{Introduction}

In two previous papers \cite{Frisch2001-1,Frisch2001-2} a grand
canonical partition function for non-interacting polymer chains was
introduced in order to compute the chain segment density in the
context of an ordered monomer
ensemble.  Here we introduce an angular dependence
between polymer segments in this formalism.  In particular, we illustrate
computations relating to the stiffness of such a chain in constraining 
geometries.
Interest in the effects of stiffness on such polymers extends from
biopolymers to liquid crystalline behavior as well as synthetic stiff
and short polyamides
\cite{Kroy96-1,Wilhelm96-1,Aharoni89-1,MuellerNedebock95-1,Liverpool98-1,Golestanian2000-1,FloryBook}.

In this paper we present three different models of confined 
chains with stiffness.  These all involve specific
formulations of the stiffness and nature of the chains in the monomer
ensemble. In one case analytic expressions for a polymer
on a cubic lattice can be obtained.  For chains in a continuum
we show that torsional and flexural rigidities can be
incorporated naturally by the monomer ensemble.  In a spherical
confining region it is possible to solve the associated integral
equations numerically.

We use the previously introduced concepts
\cite{Frisch2001-1,Frisch2001-2}.  By characterizing a bond position by a
vector $\bm{r}$ specifying its geometrical center and orientation, 
a bond fugacity $z(\bm{r})=\left\langle \bm{r} \left|
    z \right| \bm{r} \right\rangle$ and interaction weight
$w(\bm{r},\bm{r}')= \left\langle \bm{r} \left|
    w \right| \bm{r}' \right\rangle$, i.e. a Boltzmann factor, can be defined.  The
partition function for $N$ bonds 
\begin{eqnarray}
\Xi_N & = &  \int \left[ d\bm{r}_1\,d\bm{r}_2 \ldots d\bm{r}_N \right]\,\, 
\left\langle \bm{r}_1 \left| z \right| \bm{r}_1 \right\rangle
\left\langle \bm{r}_1 \left| w \right| \bm{r}_2 \right\rangle \ldots
\nonumber \\
& &
\ldots \times
\left\langle \bm{r}_{N-1} \left| w \right| \bm{r}_N \right\rangle 
\left\langle \bm{r}_N \left| z \right| \bm{r}_N \right\rangle.
\end{eqnarray}
This is used  
in order to write the expression for the grand canonical partition
function as follows:
\begin{eqnarray}
  \Xi & = & 1 + \sum_{N=1}^\infty \left\langle \bm{1} \left| z \left( wz
    \right)^{N-1} \right| \bm{1} \right\rangle \nonumber \\
& = &1+  \left\langle \bm{1} \left| z \left( I - wz \right)^{-1} \right| \bm{1}
  \right\rangle.
\label{Eq:GrandPart}
\end{eqnarray}
The vector $\left| \bm{1} \right\rangle$ is the vector of ones,
resulting in the sum/integral
over all spatial (and angular) locations.
One can write for the number density:
\begin{equation}
n(r) = \frac{1}{\Xi} z(\bm{r}) \frac{\delta \Xi}{\delta z(\bm{r})}
\end{equation}
which gives
\begin{equation}
n(\bm{r}) = \frac{1}{\Xi} \left\langle \bm{1} \left| \left( I - zw
    \right)^{-1} \right| \bm{r} \right\rangle z(\bm{r}) \left\langle \bm{r} \left|
    \left( I - wz \right)^{-1} \right| \bm{1} \right\rangle.
\end{equation}
By defining $\psi$ and $\hat{\psi}$
\begin{subequations}
\begin{eqnarray}
\psi(\bm{r}) & = & \left\langle \bm{r} \left| \left(I-wz\right)^{-1}
  \right| \bm{1} \right\rangle,
\label{Eq:PsiDef1}
\\
\hat{\psi}(\bm{r}) & = & \left\langle \bm{1} \left| \left(I-zw\right)^{-1}
  \right| \bm{r} \right\rangle,
\label{Eq:PsiDef2}
\end{eqnarray}
\end{subequations}
one can simplify the expression for the density:
\begin{equation}
\frac{n(\bm{r})}{z(\bm{r})} = \frac{\psi (\bm{r})\hat{\psi}(\bm{r})}{\Xi}
\label{Eq:DensityExpr}.
\end{equation}
The grand canonical partition function is calculated from equation
(\ref{Eq:GrandPart}) as
\begin{equation}
\Xi = 1 + \int d\bm{r} \,\, z(\bm{r}) \psi(\bm{r}).
\end{equation}
The average degree of polymerization is given by
\begin{equation}
\xi = \int d\bm{r}\,\, n(\bm{r}).
\label{Eq:AvPolDeg}
\end{equation}
The solution of expressions~(\ref{Eq:PsiDef1}) and (\ref{Eq:PsiDef2}) 
plays the central role in 
our calculations of the density.

Whereas in the previous works~\cite{Frisch2001-1,Frisch2001-2} the
physical interpretation of the vector $\bm{r}$ represented the location
of the junction between any two segments of the polymer chains, the
formalism is identical when a larger degree of freedom is represented
by a vector of such a kind.

The different approaches in this paper involve specific
formulations of the stiffness which are incorporated into equations
(\ref{Eq:PsiDef1}) and (\ref{Eq:PsiDef2}) for the functions $\psi$ 
in order to determine an expression for the density function
(\ref{Eq:DensityExpr}).  Section \ref{Sec:Lattice} shows the
computation for the
density of a discrete polymer confined between two parallel plates.
In the subsequent Section \ref{Sec:IntEq} a general form for
the integral equations for $\psi$ of a chain with bending rigidity and
torsion is derived.  This method is illustrated by numerical
results for the solution of the integral equations for a chain in a
spherical container.  Finally, a simple example for a differential
equation formalism is presented in Section \ref{Sec:DiffEq}.


\section{Polymer on a cubic lattice\label{Sec:Lattice}}

We make use of a discrete formalism for stiff chains.  (A similar
model was developed in \cite{MuellerNedebock2001-1}.  It was treated
in a canonical ensemble without any confinement .)
In addition to associating a position on a cubic lattice, each segment 
also has one of six possible directions along the lattice.
The bond direction is added to the
previous bond position to give the next bond position for the chain.  
We map the states for the directions onto real-space unit vectors 
\begin{equation}
\left\langle \sigma \right| \in \left\{ \left\langle 1 \right|,
\ldots \left\langle 6 \right| \right\} \leftrightarrow 
\left\{ {\hat{x}}, {\hat{y}}, {\hat{z}}, -{\hat{x}},
-{\hat{y}}, -{\hat{z}}\right\} = \left\{ {\hat{t}}_\sigma
  \right\}
\end{equation}
in a cartoon representation in which all bonds lie along coordinate axes
and assign to each pair of bonds the weight
\begin{eqnarray}
w( \bm{r}_1, \sigma_1; \bm{r}_2, \sigma_2 ) & = & 
\delta\left[
\bm{r}_2 - \left( \bm{r}_1 + \frac{1}{2} {\hat{t}}_{\sigma_1} +
  \frac{1}{2} {\hat{t}}_{\sigma_2} \right) \right]
\nonumber \\
& & \times 
\left\{
\begin{array}{ll}
1 & \text{, if }\hat{t}_{\sigma_1} \cdot  \hat{t}_{\sigma_2}=1 \\
a & \text{, if } \hat{t}_{\sigma_1} \cdot \hat{t}_{\sigma_2}=0 \\
b & \text{, if } \hat{t}_{\sigma_1} \cdot \hat{t}_{\sigma_2}=-1
\end{array}
\right.
\label{Eq:PottsW}
\end{eqnarray}
The first factor in the expression above constrains the segments of
the polymer; the second factor is responsible for the bending
energetics.  We express all lengths in terms of the step-length of the 
walk.
Due to asymmetry under exchange of $\bm{r}_1$ and $\bm{r}_2$ of this
interaction, two functions $\psi$, as defined earlier in
equations~(\ref{Eq:PsiDef1}) and (\ref{Eq:PsiDef2}), are invoked:
\begin{subequations}
\begin{eqnarray}
\hat{\psi} ( \bm{r},\sigma) & = & \left\langle \bm{1} \left| \left( I
      - zw \right)^{-1} \right| \bm{r},\sigma \right\rangle \\
\psi  ( \bm{r},\sigma) & = & \left\langle \bm{r}, \sigma \left| \left( I
      - wz \right)^{-1} \right| \bm{1} \right\rangle 
\end{eqnarray}
\end{subequations}
such that
\begin{subequations}
\begin{eqnarray}
1 & = & 
- \sum_{\sigma'}\int d^3 r'\,\, \left\langle
  \bm{r},\sigma \left| w^T \right| \bm{r}', \sigma' \right\rangle
z(\bm{r}',\sigma') \hat{\psi}(\bm{r}', \sigma')
\nonumber \\
& & + \hat{\psi}(\bm{r},\sigma) 
\label{Eq:PsiCondition1}
\\
1 & = & - \sum_{\sigma'}\int d^3 r'\,\, \left\langle
  \bm{r},\sigma \left| w \right| \bm{r}', \sigma' \right\rangle
z(\bm{r}',\sigma') {\psi}(\bm{r}', \sigma')
\nonumber \\
& & + {\psi}(\bm{r},\sigma)
\label{Eq:PsiCondition2}
\end{eqnarray}
\end{subequations}
where $d^3\,r'$ is here a delta--function measure that converts
integrals to sums over half--space lattices.

We investigate this polymer located between two parallel
plates located at $r_z=\pm r_z^0=\pm N$.
\textit{At the plates} the boundary conditions require that a segment
of the polymer be
oriented in parallel to the plate or perpendicularly away from it,
but not perpendicularly into it:
\begin{equation}
z(r_z, \sigma) = \left\{ \begin{array}{ll}
0, & |r_z|>r_z^0\\
0, & r_z=r_z^0\text{ and }\sigma=3\\
0, & r_z=-r_z^0\text{ and }\sigma=6\\
z_0, & \text{otherwise}
\end{array} \right.
\end{equation}
Due to symmetry the functions $\psi$ and $\hat{\psi}$ depend only on
the $z$--component of position and on $\sigma$.  
The $x$ and $y$ components are confined to a fixed large length and
all thermodynamic potentials normalized appropriately.
Furthermore, by comparing equations
(\ref{Eq:PsiCondition1}) and (\ref{Eq:PsiCondition2}) for $\psi$ and
$\hat{\psi}$ with the weight
(\ref{Eq:PottsW}) substituted, we conclude that
\begin{equation}
\hat{\psi} \left(r_z, \sigma \right) = \psi\left( r_z, (\sigma+3)
  \text{ mod }6\right),
\end{equation}
with the convention that $\hat{\psi}(r_z,6)=\hat{\psi}(r_z,0)$.
Under the abovementioned conditions we introduce the convenient notation:
\begin{equation}
\psi \left( r_z, \sigma \right) = \left\{ 
\begin{array}{ll}
\psi_\parallel \left( r_z \right), & \sigma=1,2,4,5\\
\psi_{\uparrow} \left( r_z \right), & \sigma=3 \\
\psi_{\downarrow} \left( r_z \right), & \sigma=6
\end{array}
\right.  .
\label{Eq:PsiDirectionsDef}
\end{equation} 
Symmetry dictates that
\begin{subequations}
\begin{eqnarray}
\psi_\parallel \left( r_z \right) & = & \psi_\parallel \left( -r_z
\right) \text{ and}
\label{Eq:Symm1}
\\
\psi_{\uparrow} \left( r_z \right) & = & \psi_{\downarrow} \left( -r_z
\right).
\label{Eq:Symm2}
\end{eqnarray}
\end{subequations}
Since $\psi$ is defined in half lattice constants through the weight
(\ref{Eq:PottsW}) we shall also use the notation $\psi_\uparrow(r_z) =
\psi_{\uparrow, m}$, where $m$ is an integer or half-integer, etc.,
interchangeably.

The following two subsections \ref{Subsec:PottsEqSys} and
\ref{Subsec:BoundaryPotts} contain the information required to solve
the equations (\ref{Eq:PsiCondition1}) and (\ref{Eq:PsiCondition2})
for $\psi$ and $\hat{\psi}$.  The reader not interested in the
procedure for solution may skip to subsection
\ref{Subsec:DiscreteResults}.

\subsection{System of equations\label{Subsec:PottsEqSys}}

By inserting the bending energy and chain position, $-r_z^0 +1 \leq
r_z \leq  r_z^0 -1,$ the factor $w$ of
equation (\ref{Eq:PottsW}) into the condition for $\psi$, equation
(\ref{Eq:PsiCondition2}), the following equations are obtained after 
suitable translations:
\begin{subequations}
\begin{eqnarray}
1 & = & \psi_\parallel\left(r_z\right) \left[1-z_0\left(
1 + 2a +b \right)
 \right] -
az_0 \psi_{\uparrow}\left(r_z+{\textstyle\frac{1}{2}}\right)
\nonumber \\
& &  - az_0
\psi_{\downarrow}\left(r_z- {\textstyle\frac{1}{2}}\right)
\label{Eq:PottsSys1}
\\
1 & = & \psi_{\uparrow}\left( r_z - {\textstyle\frac{1}{2}} \right) - bz_0
\psi_{\downarrow}\left( r_z - {\textstyle\frac{1}{2}} \right) 
\nonumber \\
& & - z_0
\psi_{\uparrow}\left( r_z + {\textstyle\frac{1}{2}} \right) - 4az_0
\psi_\parallel \left( r_z \right) 
\label{Eq:PottsSys2}
\\
1 & = & \psi_{\downarrow}\left(r_z+{\textstyle\frac{1}{2}} \right) -
bz_0\psi_{\uparrow} \left( r_z + {\textstyle\frac{1}{2}} \right) 
\nonumber \\
 & & 
-z_0 \psi_{\downarrow} \left( r_z - {\textstyle\frac{1}{2}} \right) - 4az_0
\psi_\parallel \left(r_z \right).
\label{Eq:PottsSys3}
\end{eqnarray}
\end{subequations}
These equations are valid away from the two plates acting as
boundaries to the system.  Consequently we shall refer to calculations 
relating to the above conditions as those pertaining to the ``bulk.''
By using the expressions from the above system (equations
(\ref{Eq:PottsSys1}) -- (\ref{Eq:PottsSys3}))
$\psi_{\parallel}(r_z)$ can be eliminated, leaving 
equations expressing $\psi_{\uparrow}(r_z+1/2)$ and
$\psi_{\downarrow}(r_z+1/2)$ in terms of 
$\psi_{\downarrow}(r_z - 1/2)$ and $\psi_{\uparrow}(r_z-1/2)$.  By
defining the column vector
\begin{equation}
\bm{\psi}(r) = \left( \begin{array}{c} \psi_\uparrow (r) \\
\psi_\downarrow (r) \end{array} \right)
\end{equation}
it is possible to relate functions of $\psi$ at different steps
by
\begin{equation}
\bm{\psi}(r+1/2) = \bm{C}\cdot \bm{\psi}(r-1/2) + \bm{D}
\label{Eq:LadderBasic}
\end{equation}
with the matrices
\begin{subequations}
\begin{eqnarray}
\bm{C} & = & \left( \begin{array}{cc}
C_1 & C_2
\\
C_3 & C_4
\end{array} \right) \\
\bm{D} & = & \left( \begin{array}{c}
D_1
\\
-z_0(1-b)D_1
\end{array} \right)\\
\text{with} & & \nonumber \\
C_1 & = & \frac{1-z_0(1+2a+b)}{z_0\left[ 1- z_0(1+2a+b) + 4a^2z_0
  \right]}
\\
& = & \frac{ \left( 1 + C_2 \right)}{z_0(1-b)}
\\
C_2 & = & -\frac{b\left(
1-z_0(1+2a+b)\right) +4a^2 z_0
}{\left[ 1- z_0(1+2a+b)+4a^2 z_0\right]}
\\
C_3 & = & -C_2
\\
C_4 
& = & z_0(1-b)\left( 1 - C_2 \right)
\\
D_1 & = & -\frac{1-z_0(1+2a+b)+4az_0}{z_0\left[ 1- z_0(1+2a+b)+4a^2 z_0\right]}
\end{eqnarray}
\end{subequations}  

As a consequence any ``bulk'' values of $\bm{\psi}$ can be calculated
given the values of $\psi$ at a point (integer and half-integer)
on the lattice:
\begin{subequations}
\begin{eqnarray} 
\bm{\psi}_{n+m} & = & \bm{C}^n \bm{\psi}_m  
\nonumber \\
& & +\left( \bm{C} - \openone
\right)^{-1} \left( \bm{C}^n - \openone \right) \bm{D} 
\label{Eq:Ladder0}\\
\bm{\psi}_{n+m+1/2} & = & \bm{C}^n \bm{\psi}_{m+1/2}  
\nonumber \\
& & +
\left( \bm{C} - \openone
\right)^{-1} \left( \bm{C}^n - \openone \right) \bm{D} 
\label{Eq:LadderHalf}
\end{eqnarray}
\end{subequations}
Before commencing on further 
calculations we note:
\begin{itemize}
\item
According to equation (\ref{Eq:PottsSys1}) $\psi_\parallel (r)$ can be 
computed with the knowledge of $\bm{\psi} (r \pm 1/2)$.
\item
The matrix $\bm{C}$ can be written in terms of $C_2$ as follows:
\begin{equation}
\bm{C} = \left( \begin{array}{cc}
\frac{1+C_2}{z_0(1-b)} & C_2 \\
-C_2 & z_0(1-b) (1-C_2)
\end{array} \right).
\end{equation}
A simple calculation shows that the determinant of the matrix is 1,
which means that its two eigenvalues are inverses of one another.
\end{itemize}

\subsection{Boundary conditions and solution\label{Subsec:BoundaryPotts}}

To determine values
of $\bm{\psi}$ it is necessary to use the equations (\ref{Eq:Ladder0}) 
and (\ref{Eq:LadderHalf}) in conjunction with the conditions at the
plates confining the polymer.
When $r_z \geq N-{\textstyle\frac{1}{2}}$ it
is necessary to refer to the full equations for $\psi$
(\ref{Eq:PsiCondition2}) rather than the bulk values used
in the preceding subsection.  Equations at the upper plate, for
example, are readily 
derived and recorded in Appendix \ref{App:PsiBoundaries}.

Equations (\ref{Eq:A4b}) and (\ref{Eq:A5}) relate $\psi_{\uparrow N-1}$ to
$\psi_{\downarrow N-1}$ and equations (\ref{Eq:A3c}) and
(\ref{Eq:A4a}) relate $\psi_{\uparrow N-1/2}$ to
$\psi_{\downarrow N-1/2}$.
\begin{subequations}
\begin{eqnarray}
1 & = & -\frac{4az_0}{1-z_0(1+2a+b)} + \psi_{\uparrow, N-1} 
\label{Eq:BC1}
 \\
& & - \left( \frac{4a^2 z_0^2}{1-z_0(1+2a+b)} + bz_0 \right) \psi_{\downarrow,
  N-1}\nonumber
\\
1 & = & -\frac{4az_0}{1-z_0(1+2a+b)} + \psi_{\uparrow, N-1/2} 
\label{Eq:BC2}
 \\
& & - \left( \frac{4a^2 z_0^2}{1-z_0(1+2a+b)} + bz_0 \right) \psi_{\downarrow,
  N-1/2}\nonumber
\end{eqnarray}
Similarly, one can derive for the bottom plate:
\begin{eqnarray}
1 & = & -\frac{4az_0}{1-z_0(1+2a+b)} + \psi_{\downarrow, -N+1} 
\\
& & - \left( \frac{4a^2 z_0^2}{1-z_0(1+2a+b)} + bz_0 \right) \psi_{\uparrow,
  -N+1}\nonumber
\\
1 & = & -\frac{4az_0}{1-z_0(1+2a+b)} + \psi_{\downarrow, -N+1/2} 
\\
& & - \left( \frac{4a^2 z_0^2}{1-z_0(1+2a+b)} + bz_0 \right) \psi_{\uparrow,
  -N+1/2}\nonumber
\end{eqnarray}
\end{subequations}

A solvable system of equations now remains.  We know from the boundary 
conditions given above that the two component of the column vector
$\bm{\psi}_{N-1}$ are not independent and we also know from
equation (\ref{Eq:Ladder0}) that $\bm{\psi}_{N-1}$ is related to $\bm{\psi}_{-N+1}$.
All remaining values for the
function $\psi$ can be determined from the expressions in appendix
\ref{App:PsiBoundaries} and from the results of the preceding subsection.

The relationship between $\bm{\psi}_{N-1}$ and $\bm{\psi}_{-N+1}$ is
\begin{equation}
\bm{\psi}_{N-1} = \bm{C}^{2N-2} \bm{\psi}_{-N+1} + \left( \bm{C} -
  \openone \right)^{-1} \left( \bm{C}^{2N-2} - \openone \right)
\bm{D}.
\label{Eq:PsiRelation}
\end{equation}
Together with the boundary conditions (\ref{Eq:BC1}) and
(\ref{Eq:BC2}) it is straightforward to determine the value of
$\bm{\psi}_{N-1}$ from which all other values of $\psi$ can be
calculated.

\begin{widetext}
The left and right matrices, $\bm{L}$ and $\bm{R}$, of $\bm{C}$
diagonalize $\bm{C}$:
\begin{eqnarray}
\bm{L} \bm{C} \bm{R} & = & \left( \begin{array}{cc}
\lambda_+ & 0 \\ 0 & \lambda_- \end{array} \right) 
\equiv \left( \begin{array}{cc}
\lambda & 0 \\ 0 & \lambda^{-1} \end{array} \right)
\\
\text{where }
\bm{L} \bm{R} & = & \bm{R} \bm{L} = \openone\\
\text{with }
\bm{L} & \equiv & \left( \begin{array}{cc}
L_1 & L_2 \\ L_3 & L_4 \end{array} \right)
\\
\text{and }
\lambda_\pm & = & 
\frac{1}{2} \left[ z_0 (1-b)(1-C_2) + \frac{1+C_2}{z_0(1-b)} \right]
\nonumber  \\
& & 
\pm \frac{1}{2} \sqrt{
\left( z_0(1-b)(1-C_2) + \frac{1+C_2}{z_0(1-b)}\right)^2 -4
}
\end{eqnarray}
Equation (\ref{Eq:PsiRelation}) becomes:
\begin{eqnarray}
\bm{L} \bm{\psi}_{N-1} & = & \left( \begin{array}{cc}
\lambda_+^{2N-2} & 0 \\ 0 & \lambda_-^{2N-2} \end{array} \right)
\bm{L} \bm{\psi}_{-N+1} 
+ \left( \begin{array}{cc}
\frac{\lambda_+^{2N-2} -1}{\lambda_+ -1 } & 0 \\
0 & \frac{\lambda_-^{2N-2} -1}{\lambda_- -1 } \end{array}\right)
\bm{L} \bm{D}.
\end{eqnarray}
The solution is:
\begin{eqnarray}
\psi_{\downarrow, N-1} & = & \left\{\left[
L_3X_1-\lambda^{-2N+2}L_4X_1- \frac{\lambda^{-2N+2}-1}{\lambda^{-1}-1} 
\left(L_3 D_1+L_4 D_2 \right)
\right] \right. \nonumber \\
& & \times \left( \lambda^{-2N+2}L_3 + \lambda^{-2N+2}L_4
  X_3\right)^{-1}
\left( \lambda^{2N-2}L_1 + \lambda^{2N-2} L_2 X_2 \right)
\nonumber \\
& & \left. + \left[ \lambda^{2N-2} L_2 X_1 - L_1 X_1 +
    \frac{\lambda^{2N-2}-1}{\lambda -1} \left( L_1 D_1 + L_2 D_2
    \right) \right] \right\} /
\nonumber \\
& & \left\{ L_1 X_2 + L_2 - \lambda^{4N-4}\left(L_1 + L_2 X_2 \right)
  \left( L_3 + L_4 X_2\right)^{-1} \left( L_3 X_2 + L_4 \right)
\right\}
\label{Eq:LatticeSolution}
\end{eqnarray}
where we have defined
\begin{eqnarray}
X_1 & = & 1 + \frac{4az_0}{1-z_0(1+2a+b)}\text{ and}
\\
X_2 & = & \frac{4a^2 z_0^2}{1-z_0(1+2a+b)} + bz_0.
\end{eqnarray}
(They are the constant and coefficient in the boundary condition equations.)
Standard methods can be used to find complete (albeit lengthy)
expressions for $L_1,\ldots, L_4$.

Clearly, the solution (\ref{Eq:LatticeSolution}) is lengthy to write
out in full, although it is given explicitly.  It is simpler to
consider limiting expressions.  For the purposes of this we shall
choose a specific case where $b=a^2$ and $0<a<1$, and expand to first
order in $\epsilon=a$, i.e. the case of an extremely stiff polymer.
Functions for this scenario are labeled by a superscript ``S''.

For the stiff case and for $n \in \mathbb{Z}^+$ we have:
\begin{eqnarray}
\bm{C}^n & \simeq & \left( \begin{array}{cc}
z_0^{-n} & 0 \\ 0 & z_0^n \end{array}\right) + \mathcal{O}(\epsilon^2)
\\
\bm{E}_n & \simeq & \left( \begin{array}{cc}
\frac{z_0}{1-z_0}\left( z_0^{-n}-1\right)
 & 0 \\ 0 &
\frac{1}{z_0 -1} \left( z_0^n -1 \right)
\end{array}\right)
\times \left( \begin{array}{c}
-\frac{1}{z_0} - \frac{4}{1-z_0}\epsilon \\
1 + \frac{4z_0}{1-z_0}\epsilon \end{array} \right)
 + \mathcal{O}(\epsilon^2)
\\
\text{where } \bm{E}_n & \equiv &
\left( \bm{C} - \openone \right)^{-1} \left( \bm{C}^n - \openone \right) 
\bm{D}
\\
\psi_{\downarrow, N-1}^{\text{S}}=\psi_{\uparrow,-N+1}^{\text{S}} & \simeq &
z_0^{2N-2} + \frac{1}{z_0-1}\left(z_0^{2N-2}-1\right) + 
\epsilon\left( \frac{4z_0}{1-z_0}\right) \left[
  z_0^{2N-2}+\frac{z_0^{2N-2} -1}{z_0 -1} \right].
\\
\psi_{\downarrow,0}^{\text{S}}=\psi_{\uparrow,0}^{\text{S}} & \simeq &
z_0^{N-1}\left(1-\frac{1}{1-z_0}\right)+ \frac{1}{1-z_0} +
\epsilon \frac{4z_0}{1-z_0}\left[ \frac{1}{1-z_0} + z_0^{N-1}\left(1 -
\frac{1}{1-z_0} \right)\right]
\label{Eq:Psi0S}
\\
\psi^{\text{S}}_{\parallel, 0} & = & 
\frac{1}{1-z_0} + \epsilon \frac{2z_0}{1-z_0} \left[
\frac{1}{1-z_0} + z_0^{N-1} + \frac{1}{1-z_0}\left(1-z_0^{N-1}\right)\right]
\label{Eq:PsiP0S}
\end{eqnarray}
\end{widetext}

Similarly we have the following approximation for the floppy (``F'')
case where $a=b=1$:
\begin{equation}
\bm{C}^{\text{F}} = \left( \begin{array}{cc}
\frac{1-4z_0}{z_0} & -1 \\ 1 & 0 \end{array} \right).
\end{equation}

\subsection{The grand canonical partition function, the density, and average
  degree of polymerization\label{Subsec:DiscreteResults}} 

The grand canonical partition function, which also features in the
expression for the density (\ref{Eq:DensityExpr}), for the discrete model is given 
as usual by:
\begin{eqnarray}
\Xi & = & 1 + \sum_\sigma \sum_{\{\bm{r}\}}\,
z(\bm{r},\sigma)\psi(\bm{r},\sigma)
\nonumber \\
& = & 1 + 2z_0 \left( 4 \psi_{\parallel, -N} + \psi_{\uparrow, -N}
\right)
\\
& &
+ z_0 \left( 1, 1 \right) \cdot 
\left(\openone - \bm{C}\right)^{-1} \left\{\left(\openone -
  \bm{C}^{2N}\right) \bm{\psi}_{-N+1/2}
\right. \nonumber \\
& & \left. +  \left(\openone -
  \bm{C}^{2N-1}\right) \bm{\psi}_{-N+1}
\right. \nonumber \\
& & +   \left[
\left(\openone - \bm{C}\right)^{-1} \left(\openone -
  \bm{C}^{2N}\right) - 2N \openone \right]\cdot\bm{D}
\nonumber \\
& & + \left. \left[
\left(\openone - \bm{C}\right)^{-1} \left(\openone -
  \bm{C}^{2N-1}\right) - (2N-1) \openone \right]\cdot\bm{D}
\right\} \nonumber 
\end{eqnarray}
It is a number dependent on $N$, $z_0$, $a$, and $b$.  

The parameters of the present model can be understood to give
two different types of behavior when considering, for example, the
parallel and perpendicular orientations at the center of
the two plates.  We wish to investigate the ratio of the probability
that the segments in the middle of the plates have a perpendicular
orientation with respect to the plates to the probability that they 
are parallel to the plates
using equation~(\ref{Eq:DensityExpr})
\begin{equation}
\frac{n(0,\uparrow)}{n(0,\parallel)} =
\frac{\psi^2_{0}}{\psi_\parallel^2 (0)}
\label{Eq:RatioExpr}
\end{equation}
The values of $\psi_\uparrow(0)$ and $\psi_\parallel (0)$ at the
center are easily calculated according to the scheme in
Appendix~\ref{App:PsiBoundaries} and in preceding sections of the paper. 

\begin{widetext}
We calculate the ratio ${n(0,\uparrow)}/{n(0,\parallel)}$ for the
limiting case where $a$ and $b\equiv a^2$ are almost equal to 0 using
equations (\ref{Eq:Psi0S}) and (\ref{Eq:PsiP0S})
(the ``stiff'' case 
labeled by ``S''). 
\begin{equation}
\frac{n^{\text{S}}(0,\uparrow)}{n^{\text{S}}(0,\parallel)} \simeq  
\frac{z_0^{N-1}\left(1-\frac{1}{1-z_0}\right)+ \frac{1}{1-z_0} +
\epsilon \frac{4z_0}{1-z_0}\left[ \frac{1}{1-z_0} + z_0^{N-1}\left(1 -
\frac{1}{1-z_0} \right)\right]}{
\frac{1}{1-z_0} + \epsilon \frac{2z_0}{1-z_0} \left[
\frac{1}{1-z_0} + z_0^{N-1} + \frac{1}{1-z_0}\left(1-z_0^{N-1}\right)\right]
}
\end{equation}
For chains with a low fugacity and plates which are far apart, this
ratio approaches 1, indicating that the chain segments are isotropic
in the center.
For the stiff case it can be seen clearly that for long chains and
small plate spacing most of the polymer is parallel to the plates.
\end{widetext}

The degree of polymerization (\ref{Eq:AvPolDeg}) can also be evaluated 
\begin{eqnarray}
\xi & = & \frac{2z_0}{\Xi} \left\{ 
\left[\psi_0^2 + 2\psi_\parallel^2 \right]
\right.
\nonumber \\
& & 
\left.
+2 \sum_{i=1}^{N} \left( \psi_{\uparrow,i}\psi_{\downarrow,i} 
+ 2
  \psi_{\parallel,i}^2 \right)\right\}.
\end{eqnarray}
The summation above also contains the non--bulk contributions at the
edges of the system.

The Potts-type model which has been discussed in the present section
can also be investigated from the viewpoint of a set of differential
equations.  The discrete ``bulk equations'' can be converted into
differential equations by expanding around $r_z$ to second order.
The coupled set of differential equations can be solved by Laplace
transformation.  The need to solve for the roots of a fourth--order
polynomial for the inverse Laplace transform means that this method
does not bring about much of a simplification of the system.


\section{Integral equation\label{Sec:IntEq}}

\begin{figure}
\includegraphics[width=8cm]{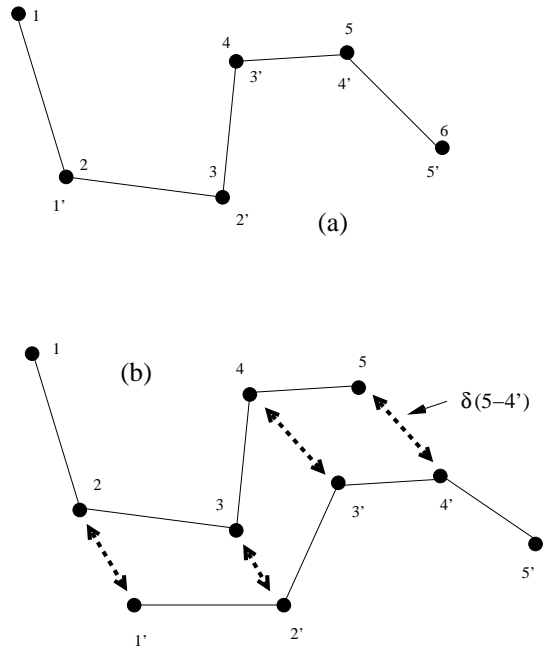}
\caption{\label{Fig:Labeling}The figure depicts the double labeling
  scheme in part (a).  This scheme can also bee seen to emerge
  from $\delta$-function Boltzmann factors connecting ``monomers'' with
  five-segment structures, as depicted in  part (b) of the figure.
  The dashed double arrows indicate which parts of the ``monomer''
  12345 must be the same physical locations on the ``monomer''
  $1'2'3'4'5'$.  In this way the dashed double arrows are delta
  functions of the respective positions.  We show one such delta
  function joining points 5 and 4'.}
\end{figure}

To obtain an integral equation for a system with stiffness one can
employ a system of double labeling of successive bonds.
Thus one can write for the partition function
\begin{equation}
\Xi_N = \int da\,da'\,db\,db'\, \ldots \left\langle a \left| w \right| a' \right\rangle
\left\langle a' \left| z \right| b \right\rangle \left\langle b \left| 
    w \right| b' \right\rangle \ldots
\end{equation}
where the $a, a', b, b', \ldots$ denote successive bonds.  The scheme
is illustrated in Figure~\ref{Fig:Labeling}.
For the bond labeled either 3 or 2' in Figure~\ref{Fig:Labeling} the
unit vector for the bend can be computed by means of
$\widehat{(\bm{r}_2 - \bm{r}_3)}$
and $\widehat{(\bm{r}_3 - \bm{r}_4)}$.  The torsional angle must be
computed by taking more vectors into account and can be constructed by
investigating, for example, $\left( \bm{r}_1- \bm{r}_2\right)\times
\left( \bm{r}_2 - \bm{r}_3 \right)$ in relation to $\left( \bm{r}_3-
  \bm{r}_4\right) \times
\left( \bm{r}_4 - \bm{r}_5 \right)$.
The full torsional and bending energy at position 3 in 
Figure~\ref{Fig:Labeling} is then expressed through a potential
$V(\bm{r}_1,\bm{r}_2,\bm{r}_3,\bm{r}_4, \bm{r}_5)=V\left(
  \vec{\bm{r}}_3 \right)$ which can be
written as a weight in the chemical potential factor $z=z\left(
  \vec{\bm{r}}_3 \right)=\exp \left(-\beta V\left( \vec{\bm{r}}_3
  \right)\right)$, 
with $\vec{\bm{r}}_3$ representing the
supervector centered around position 3.  The role of $w$ lies in
connecting the other points correctly for the preceding and succeeding 
bond-related angles. It does this as follows:
\begin{eqnarray*}
\lefteqn{
\ldots \left\langle \vec{\bm{r}} \left| z \right| \vec{\bm{r}}
\right\rangle \left\langle \vec{\bm{r}} \left| w \right|
  \vec{\bm{r}}\,' \right\rangle \left\langle \vec{\bm{r}}\,' \left| z
  \right| \vec{\bm{r}}\,' \right\rangle \ldots} \\
& = & 
\ldots \left\langle \vec{\bm{r}} \left| z \right| \vec{\bm{r}}
\right\rangle 
\delta \left( \bm{r}_1' - \bm{r}_2 \right) 
\delta \left( \bm{r}_2' - \bm{r}_3 \right)
\delta \left( \bm{r}_3' - \bm{r}_4 \right)
 \\
& & \times
\delta \left( \bm{r}_4' - \bm{r}_5 \right)
\left\langle \vec{\bm{r}}\,' \left| z
  \right| \vec{\bm{r}}\,' \right\rangle \ldots
\end{eqnarray*}

Part (b) of Figure~\ref{Fig:Labeling} shows how the preceding
mathematical prescription of labeling follows when the monomers are
viewed as having internal structure, with appropriate weights for
internal conformations.  The monomers interact so 
that appropriate parts of the substructure coincide.  This
fixes $z$ and $w$, respectively.  We remark that
there are several possibilities to incorporate flexural and
bending terms in the integral equation formalism; we shall illustrate
one such way.

The equation for $\psi\left(\vec{\bm{r}}\right)$ becomes
\begin{equation}
\psi \left(\vec{\bm{r}}\right) =\left\langle \vec{\bm{r}} \left|
    \left( I - wz \right)^{-1} \right| \bm{1} \right\rangle 
\end{equation}
leading to 
\begin{eqnarray}
1 & = & 
\psi \left( \bm{r}_1 \bm{r}_2 \bm{r}_3 \bm{r}_4 \bm{r}_5 \right) 
\nonumber \\
& & 
- \int d^3r_5' \, z\left(
\bm{r}_2 \bm{r}_3 \bm{r}_4 \bm{r}_5 \bm{r}_5' \right)
\psi \left( \bm{r}_2 \bm{r}_3 \bm{r}_4 \bm{r}_5 \bm{r}_5' \right).
\end{eqnarray}
By assuming simple bending without torsional effects, the description
can be simplified by making use of three consecutive position coordinates:
\begin{subequations}
\begin{eqnarray}
\psi\left( \bm{r}_1 \bm{r}_2 \bm{r}_3\right) & = & 1+
\int d^3r' \, z\left(
\bm{r}_2 \bm{r}_3 \bm{r}' \right)
\psi \left( \bm{r}_2 \bm{r}_3 \bm{r}'\right)
\label{Eq:BendingInt1}
\\
\text{and} & & \nonumber \\
\hat{\psi}\left( \bm{r}_1 \bm{r}_2 \bm{r}_3\right) & = & 1 +
\int d^3r' \, z\left(
\bm{r}' \bm{r}_1 \bm{r}_2 \right)
\hat{\psi} \left( \bm{r}' \bm{r}_1 \bm{r}_2\right).
\label{Eq:BendingInt2}
\end{eqnarray}
\end{subequations}

A model which lends itself readily to an iterative numerical solution
is that of the chain of segments of fixed length which is confined to
a spherical region.  A bending probability for two adjacent
segments labeled 12 and 23, with unit vectors ${\hat{n}}_{12}$ and
${\hat{n}}_{23}$, can be assigned
\begin{equation}
P\left( {\hat{n}}_{12}, {\hat{n}}_{23} \right)
= p \left( 1 +  {\hat{n}}_{12} \cdot {\hat{n}}_{23} \right).
\label{Eq:IntBendingProb}
\end{equation}
This causes the forward direction to be favored, with $p>0$.  
The function of
$z=z\left(\bm{r}_1,\bm{r}_2,\bm{r}_3\right)$ 
in this model is to restrict $\bm{r}_1-\bm{r}_2 = {\hat{n}}_{12}$
and $\bm{r}_2-\bm{r}_3 = {\hat{n}}_{23}$ to  unit vectors, and to 
keep the vectors for the spatial locations of bonds 
$\bm{r}_1,\bm{r}_2,$ and $\bm{r}_3$ from going
out of the confines of the sphere.  
Consequently, one can write
\begin{eqnarray}
z(\bm{r}_1,\bm{r}_2,\bm{r}_2)& = & p \left( 1 +  {\hat{n}}_{12} \cdot
  {\hat{n}}_{23} \right) \delta\left(\left|{\hat{n}}_{12}\right| 
-1\right)
\delta\left(\left| {\hat{n}}_{23}\right| -1  \right)
\nonumber \\
& & \times \vartheta\left(R- \left| \bm{r}_1 \right| \right)
\vartheta\left(R- \left| \bm{r}_2 \right| \right)
\vartheta\left(R- \left| \bm{r}_3 \right| \right)
\label{Eq:Intz}
\end{eqnarray}
where $R$ is the radius of the confining sphere and $\vartheta$ is the
Heavyside step function.

The spherical
symmetry and equations (\ref{Eq:Intz}) and (\ref{Eq:BendingInt1}) require the
following dependence 
\begin{equation}
\psi\left( \bm{r}_1 \bm{r}_2 \bm{r}_3\right) = 
\psi\left( \left| \bm{r}_2 \right|, {{n}}_{23,r} \right)
\end{equation}
where ${n}_{23,r}$ is the radial component of the unit vector
${\hat{n}}_{23}$.  

\subsection{Results\label{Subsec:IntResults}}

In a numerical
scheme to solve the equations, it is possible 
to iterate equation (\ref{Eq:BendingInt1}) at 
different values of the parameters.  By rewriting the basic integral
equation with $P$, given by (\ref{Eq:IntBendingProb}), one has
\begin{eqnarray}
\psi(\bm{r},n_r) & = & 1 + 
\nonumber \\
& & 
\int d{\hat{n}}' \,\, p (1 +
{\hat{n}}\cdot {\hat{n}}') \tilde{z}
\psi (\bm{r}+{\hat{n}},{\hat{n}}').
\label{Eq:ItEqn}
\end{eqnarray}
In this equation $\tilde{z}$ ensures that the positions of the bonds
remain within the spherical region.  We find
(App. \ref{App:NumIntScheme}) that $\psi$
can be split into a sum of two parts in successive spherical shells,
one of which is only a function of the radial distance $r$ and
another which is directly proportional to the radial component of the
unit vector $n_r$ multiplied by a function of $r$ only.
Therefore, to integrate from one spherical shell to the next we write:
\begin{equation}
\psi( r, n_r ) \equiv \phi_1(r) + n_r \phi_2 (r).
\label{Eq:IntSchemeIteration}
\end{equation} 
With this manner of splitting the function $\psi$ it is possible to
divide the system into a number of spherical shells for each of which a 
$\phi_1$ and $\phi_2$ have been defined.  Equation (\ref{Eq:ItEqn}) 
for $\psi$ can be iterated until the values converge.
The integration scheme for the different shells is discussed in
Appendix \ref{App:NumIntScheme}.

\begin{figure}
\includegraphics[width=6.3cm,angle=-90]{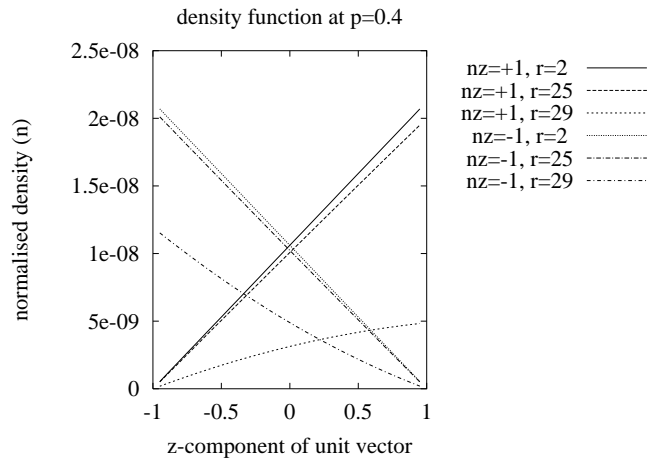}
\caption{\label{Fig:Density0.4radii}Plot of the density at different radii
  for $R=30$ and $p=0.4$.  Right and left-sloping lines represent
  forward and backward directions, respectively.}
\end{figure}

With the knowledge of $\psi$ the value of $\Xi$ can be computed, and
the density expression is:
\begin{eqnarray}
n\left({\hat{n}}_{12}, \left| \bm{r}_2 \right|, {\hat{n}}_{23}
\right) & = & \frac{1}{\Xi} p (1 +
{\hat{n}}_{12}\cdot {\hat{n}}_{23})\psi(r_2, n_{12,r})
\nonumber \\
& & \times 
\psi(r_2, n_{23,r}).
\end{eqnarray}
This five-dimensional quantity can be plotted in a variety of
manners.  In Figure \ref{Fig:Density0.4radii} we plot the density at
three different radii in dependence on the $r$--component of
${\hat{n}}_{23}$ and both unit vectors lying in the same plane.
The other directional component is chosen as lying either radially
outward $n_z=+1$ or radially inward $n_z=-1$.  For central regions of
the sphere we see that the straight configuration is favored and that 
the angular distribution is more-or-less isotropic, i.e. that both unit
vectors lying inwards-pointing or outwards--pointing is almost equally
probable.  This changes appreciably at the sides, where the radially
outward density is considerably lower than the inwards--facing case.

\begin{figure}
\includegraphics[width=6.3cm,angle=-90]{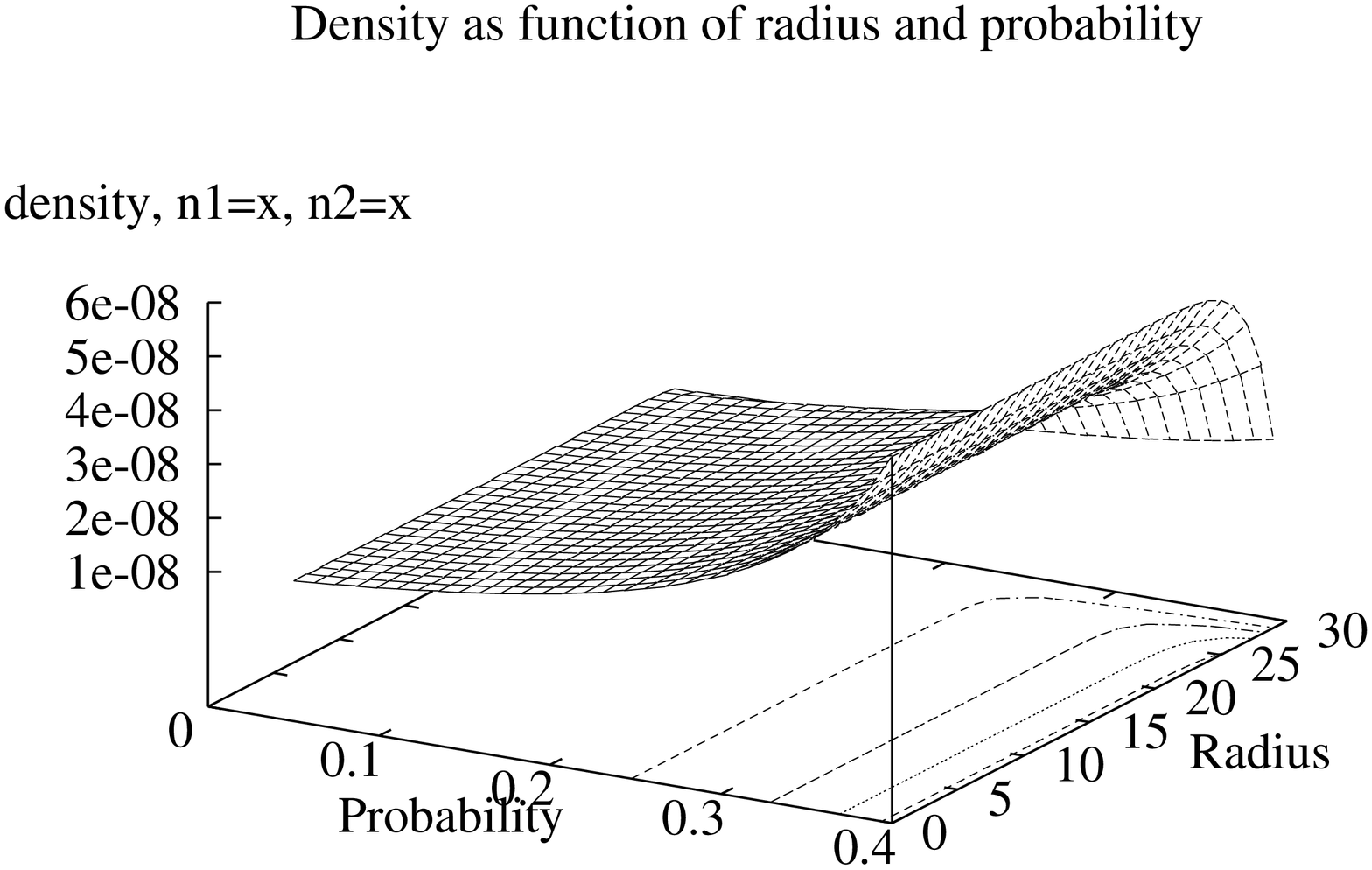}
\caption{\label{Fig:DensityRadiiProbs}Plot of the density at different radii
  for $R=30$ and and various values of $p$.}
\end{figure}

In Figure \ref{Fig:DensityRadiiProbs} the density for a both unit
vectors tangential to the radial direction and the bond being straight 
is plotted as a function of the radius and the probability $p$.  The
density decreases towards the boundary of the system, located at
$R=30$, and increases with $p$.  Note that the chemical potential is
built in through $p$.

From these graphs a clear picture emerges of a chain which is
homogeneous in the center of the confining sphere and which becomes
depleted at the boundaries.  At these boundaries a tangential
orientation of the segments is considerably favored above the
perpendicular (radial) case.

\begin{figure}
\includegraphics[width=6.3cm,angle=-90]{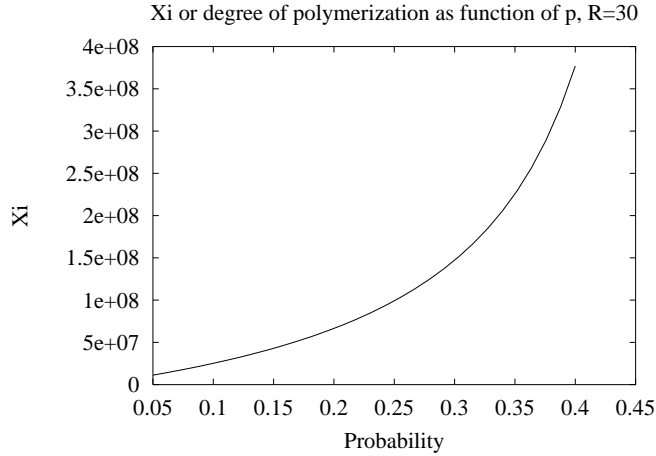}
\caption{\label{Fig:XiProbs}Plot of the degree of polymerization at a
  function of $p$.}
\end{figure}

\begin{figure}
\includegraphics[width=6.3cm,angle=-90]{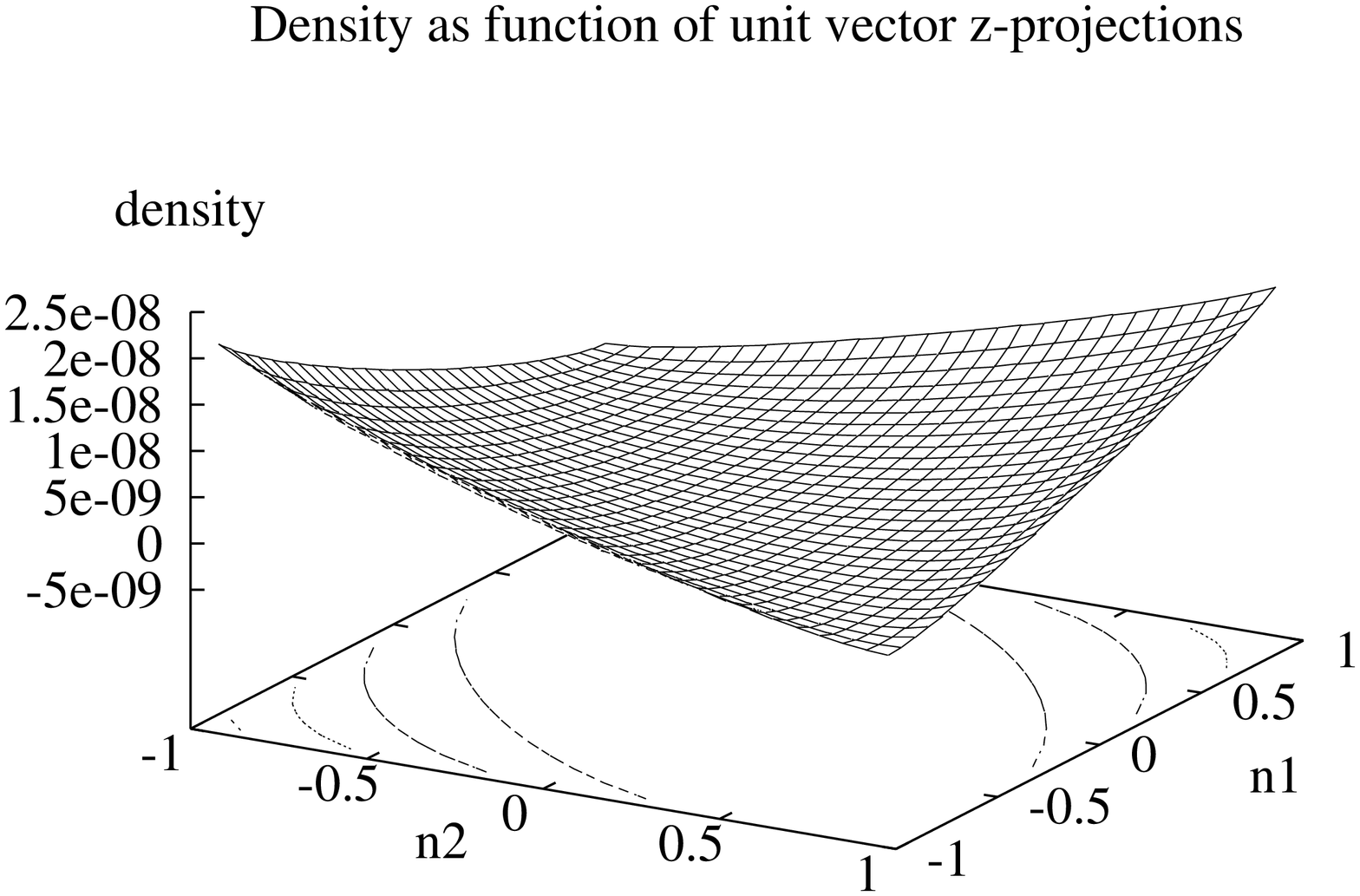}
\caption{\label{Fig:DensityNN}Density for different $z$--direction
  projections at $p=0.4$.}
\end{figure}

Figures \ref{Fig:XiProbs} and \ref{Fig:DensityNN} indicate aspects of
the normalization and dependences on orientation.  In Figure
\ref{Fig:XiProbs} the dependence of the degree of polymerization on
the combined probability and chemical potential $p$ is shown.

\subsection{Possible alternative methods for
  solution\label{Subsec:IntAlternatives}}

Another method to solve 
the integral equation is making use of an expansion in terms of 
eigenfunctions, and using successive substitutions to determine
coefficients.  For the case of equation (\ref{Eq:BendingInt1}) a weight
\begin{eqnarray}
z\left(\bm{r}_1 \bm{r}_2 \bm{r}_3 \right) & = & {\cal N}\tilde{\vartheta}
\left[
\left( \bm{r}_2 - \bm{r}_1 \right) \cdot \left( \bm{r}_3 - \bm{r}_2
\right) \right]^2
\nonumber \\
& \times & \left( \bm{r}_2 - \bm{r}_1 \right)^2 \exp \left(
  -\frac{1}{\ell^2} \left( \bm{r}_2 - \bm{r}_1 \right)^2 \right)
\nonumber \\
& \times & \left( \bm{r}_3 - \bm{r}_2 \right)^2 \exp \left(
  -\frac{1}{\ell^2} \left( \bm{r}_3 - \bm{r}_2 \right)^2 \right)
\end{eqnarray} 
could be introduced.  The first factor after $\tilde{\vartheta}$
[which, as in (\ref{Eq:Intz}), confines points to an appropriate region] 
represents the bending interaction term of
the type $\cos^2 \theta$ between two bonds, while the peaked functions 
$x^2\exp(-x^2)$ set a length scale to the segments.  ${\cal N}$ is a
normalization. 

The weight $z$ and $\psi$ can be expressed as a sum of (generalized) Hermite
polynomials~\cite{Erdelyi53-1}
\begin{eqnarray}
z\left(\bm{r}_1 \bm{r}_2 \bm{r}_3 \right) & = & \sum_{\bm{n}}
q_{\bm{n}} \left\langle 
\left.  H_{\bm{n}} \right| 
\bm{r}_1 \bm{r}_2 \bm{r}_3  \right\rangle\\
\psi\left(\bm{r}_1 \bm{r}_2 \bm{r}_3 \right) & = & \sum_{\bm{m}}
y_{\bm{m}} \left\langle
 \bm{r}_1 \bm{r}_2 \bm{r}_3 \left|
  H_{\bm{m}} \right. \right\rangle.
\end{eqnarray}
These functions could be be inserted into the bending integral equation
(\ref{Eq:BendingInt1}) and terms compared.


\section{Differential equation form\label{Sec:DiffEq}}

In \cite{Frisch2001-1} the density distribution for a polymer confined 
to a spherical cavity was computed by expressing $\psi$ in
equation~(\ref{Eq:PsiDef1}) as
\begin{equation}
z\psi = w^{-1} \left(\psi -1 \right)
\end{equation}
and by making the appropriate choices for the fugacity inside and outside the 
cavity.  The inverse Boltzmann factor $w^{-1}$ was taken to be the
differential operator of which $w$ is the Green function.  A suitable
choice was that of a Yukawa form leading to a Helmholtz operator:
\begin{eqnarray}
\delta\left(\bm{r} - \bm{r}'\right) & = & w^{-1} \frac{A
  e^{-K|\bm{r}-\bm{r}'|}}{ \left| \bm{r} - \bm{r}' \right|}
\nonumber \\
 & = & 
\frac{-1}{4\pi A}\left(\nabla^2 - K^2 \right) w.
\end{eqnarray}
A solution can be found readily for this system.

In order to introduce stiffness a segment of the polymer chain is now
described by the positions of its ends $\bm{x}, \bm{x}'$ and the
orientation of those ends ${\hat{n}},{\hat{n}}'$.  Clearly, for
a rod $\left(\bm{x}- \bm{x}'\right)$, ${\hat{n}}$ and
${\hat{n}}'$ are related.  The vector $\left|\left. \bm{r}
  \right.\right\rangle$ then has the dependence $\left|\left. \bm{r}
  \right.\right\rangle = \left|\left. \bm{x}, \vartheta, \phi
  \right.\right\rangle$.
By noting that
\begin{equation}
\left(\frac{-1}{a^2}\right)\left( \frac{\partial^2}{\partial \alpha^2} -
  a^2 \right) \left[ \frac{a}{2} e^{-a\left| \alpha - \alpha' \right|}
\right] = \delta \left(\alpha - \alpha' \right)
\end{equation}
a simple multiplicative, bending ``Boltzmann'' factor is introduced:
\begin{equation}
w\left(  \bm{x}, \vartheta, \phi;  \bm{x}', \vartheta', \phi'\right)
= \frac{Aab}{4|x-x'|}e^{-a|\vartheta-\vartheta'|-b|\phi-\phi'| - K|x-x'|}.
\end{equation}
The substitution
\begin{equation}
\psi -1 = \frac{\eta(\bm{r})}{|\bm{r}|} \xi(\vartheta) \zeta(\phi)
\end{equation}
will separate the variables.  For the spherical container we need to
solve 
\begin{equation}
\left(\frac{\partial^2}{\partial \vartheta^2}-a^2\right)
\left(\frac{\partial^2}{\partial \phi^2}-b^2\right)
\left(\nabla^2-K^2\right) \eta(x)\xi(\vartheta) 
\zeta(\phi) = 0
\end{equation}
for the outside and
\begin{eqnarray}
z_0 & = & 
\left(\frac{\partial^2}{\partial \vartheta^2}-a^2\right)
\left(\frac{\partial^2}{\partial \phi^2}-b^2\right)
\left(\nabla^2-K^2\right) \frac{\eta}{x}\xi 
\zeta \nonumber \\
& & + 4\pi a^2 b^2 z_0 \frac{\eta}{x}\xi\zeta
\end{eqnarray}
for the inside.  For the outside the usual radial solution is
obtained, whereas for the inside of the sphere we have:
\begin{eqnarray}
\left(\frac{\partial^2}{\partial \vartheta^2}-a^2\right) \xi & = &
c_\xi \xi \label{Eq:XiDiff}\\
\left(\frac{\partial^2}{\partial \phi^2}-b^2\right) \zeta & = &
c_\zeta \zeta \label{Eq:ZetaDiff}\\
\left(\frac{\partial^2}{\partial x^2}-K^2\right) \eta & = & c_\eta
\eta
\end{eqnarray}
where 
\begin{equation}
c_\xi c_\zeta c_\eta = - 4\pi A a^2 b^2.\label{Eq:ProductCondition}
\end{equation}
Here the radial distance and orientation of parts of the chain are
completely decoupled.  This is physically acceptable.  
We impose the fact that the boundary conditions 
are cyclic in that $\xi(\vartheta+2\pi)=\xi(\vartheta)$ and
$\zeta(\phi+\pi)=\zeta (\phi)$.  For the polymer confined to the
spherical cavity the external solution for $r$ only is required.  The
results for $r$ are identical to the results in \cite{Frisch2001-1}:
\begin{eqnarray}
r(\psi(r)  -1 ) & = & \frac{K^2-{K'}^2}{{K'}^2} \left( r
\right. \\
& & \left. - \frac{(1+KR)\sinh K'r }{K\sinh K'R 
+ K' \cosh K' R}\right),\,r\leq R.\nonumber 
\end{eqnarray} 
Since the solutions to equations (\ref{Eq:XiDiff}) and
(\ref{Eq:ZetaDiff}) have to be periodic or constant, the condition on
the constants $c_\xi$ and $c_\zeta$ are that
\begin{eqnarray}
c_\xi + a^2 & \leq & 0 \text{ and}\\
c_\zeta + b^2 & \leq & 0.
\end{eqnarray}
With (\ref{Eq:ProductCondition}) $c_\eta$ must be negative.

\section{Conclusions}

We have demonstrated how the formalism
accommodates potentials with angular dependence in three different
ways.  The formalism is quite generally applicable to a variety of
problems with more than positional degrees of freedom.

In subsequent work we shall develop this formalism for a path integral 
formulation to include investigation of the effects of lateral
interactions. 

\section*{Acknowledgements}

HLF acknowledges the financial support of NSF Grant DMR 9628224 
and a Fulbright Fellowship.  KKMN acknowledges the financial support
of the National Research Foundation of South Africa.



\appendix

\section{Lattice boundary condition and central value
  equations\label{App:PsiBoundaries}} 

Here we employ equation (\ref{Eq:PsiCondition2}) for different values
of $r_z$ and $\sigma$ near the boundaries.
\begin{itemize}
\item For $r_z=r^0_z+1$, and for $\sigma=\uparrow, \downarrow, \text{
    and }\parallel$, respectively:
\begin{subequations}
\begin{eqnarray}
1 & = & \psi_\uparrow(r_z^0+1) \label{Eq:A1a}\\
1 & = & - z_0 \psi_\downarrow(r_z^0) + \psi_\downarrow (r_z^0+1) \label{Eq:A1b}\\
1 & = & \psi_\parallel (r_z^0+1) \label{Eq:A1c}
\end{eqnarray}
\end{subequations}
\item For $r_z=r^0_z+{\textstyle\frac{1}{2}}$, and for
  $\sigma=\uparrow, \downarrow, \text{
    and }\parallel$, respectively:
\begin{subequations}
\begin{eqnarray}
1 & = & \psi_\uparrow(r_z^0+{\textstyle\frac{1}{2}})  \label{Eq:A2a}\\
1 & = & -4az_0 \psi_\parallel(r_z^0) - z_0
\psi_\downarrow(r_z^0-{\textstyle\frac{1}{2}}) 
\nonumber \\
& & + \psi_\downarrow
(r_z^0+{\textstyle\frac{1}{2}} ) \label{Eq:A2b}\\
1 & = & -az_0 \psi_\downarrow (r_z^0) + \psi_\parallel
(r_z^0+{\textstyle\frac{1}{2}} ) \label{Eq:A2c}
\end{eqnarray}
\end{subequations}
\item For $r_z=r^0_z$, and for $\sigma=\uparrow, 
\downarrow, \text{
    and }\parallel$, respectively:
\begin{subequations}
\begin{eqnarray}
1 & = & -bz_0 \psi_\downarrow(r_z^0) + \psi_\uparrow(r_z^0)  \label{Eq:A3a}\\
1 & = & -4az_0 \psi_\parallel(r_z^0-{\textstyle\frac{1}{2}}) - z_0
\psi_\downarrow(r_z^0-1) 
\nonumber \\
& & + \psi_\downarrow
(r_z^0) \label{Eq:A3b}\\
1 & = & -az_0 \psi_\downarrow (r_z^0-{\textstyle\frac{1}{2}}) 
\nonumber \\
& &
+ 
\left[ 1 -z_0(1+2a+b)\right]
\psi_\parallel
(r_z^0 ) \label{Eq:A3c}
\end{eqnarray}
\end{subequations}
\item For $r_z=r^0_z-{\textstyle\frac{1}{2}}$, and for 
$\sigma=\uparrow
\text{
    and }\parallel$, respectively:
\begin{subequations}
\begin{eqnarray}
1 & = & -4az_0 \psi_\parallel(r_z^0) 
\nonumber \\
& & 
- bz_0\psi_\downarrow(r_z^0-{\textstyle\frac{1}{2}}) +
\psi_\uparrow(r_z^0-{\textstyle\frac{1}{2}})   \label{Eq:A4a}\\
1 & = & \left[ 1 - z_0 (1 + 2a+b)\right] \psi_\parallel (r_z^0 -
{\textstyle\frac{1}{2}}) 
\nonumber \\
& & -az_0 \psi_\downarrow (r_z^0 -1) \label{Eq:A4b}
\end{eqnarray}
\end{subequations}
\item For $r_z=r_z^0 -1$ and $\sigma=\uparrow$:
\begin{eqnarray}
1 & = & -4az_0 \psi_\parallel (r_z^0 - {\textstyle\frac{1}{2}}) 
\nonumber \\
& &
-bz_0 \psi_\downarrow (r_z^0 -1) + \psi_\uparrow (r_z^0-1)  \label{Eq:A5}
\end{eqnarray}
\end{itemize}

Halfway between the plates the symmetry dictates that
\begin{equation}
\psi_{\uparrow,0} = \psi_{\downarrow, 0}.
\end{equation}


\section{Numerical integration scheme for sphere-constrained
  walk\label{App:NumIntScheme}} 

\begin{figure}
\includegraphics[width=6cm,angle=-90]{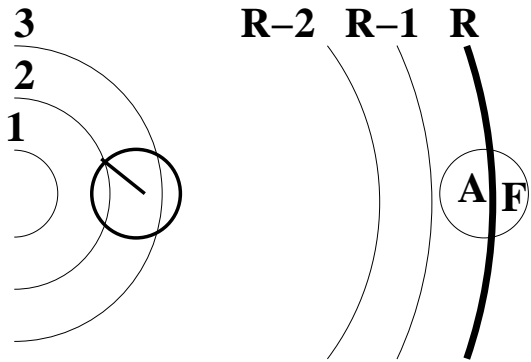}
\caption{\label{Fig:IntScheme} Schematic representation of the
  integration procedure over shells. The shell numbering is
  illustrated, as well as the mixing between different shells.  ``F''
  represents the forbidden region for any bond of the polymer.}
\end{figure}

In order to elucidate the integration scheme used for the numerical
calculations we refer to Figure \ref{Fig:IntScheme}.  The spherical
geometry of the system is shown here up to the edge of the system.
For each shell of thickness 1 we calculate $\psi_i = \phi_i^{(1)} +
n_z \phi_i^{(2)}$.
Since the shells have the thickness of the radius of the bond, at each
stage there are contributions from the two adjoining shells, according 
to equation (\ref{Eq:ItEqn}).  These can be summed if we assume that
the values of $\phi_i^{(1)}$ and 
$\phi_i^{(2)}$ are constant and approximately equal to the value of
the $\phi$'s in the \emph{middle} of each shell.  In the final shell
ending at the radius $R$, the bond vector is permitted to move only
in the allowed region A with zero weight in the forbidden region F of
Figure \ref{Fig:IntScheme}. The set of $\left\{ \phi^{(1)}_i,
  \phi^{(2)}_i \right\}$ is iterated through equation (\ref{Eq:ItEqn}) 
until the values no longer change.


\bibstyle{/home/kkmn/revtex4/apsrev.bst}

\end{document}